\documentclass{PoS}
\Preprint{HERWIG-2016-01, KA-TP-03-2016, MCnet-16-01}

\usepackage{tikz}
\newcommand*\tikzmark[2][mark]{%
  \ifmmode
    \tikz[remember picture,baseline=(#1.base)]\node(#1){$#2$};%
  \else
    \tikz[remember picture,baseline=(#1.base)]\node(#1){#2};%
  \fi
}

\DeclareMathAlphabet{\mathpzc}{OT1}{pzc}{m}{it}

\usepackage{amsmath}

\usepackage{caption}

\title{NLO efforts in Herwig++}

\ShortTitle{NLO efforts in Herwig++}

\author{%
Johannes Bellm$^{1,2}$,
Stefan Gieseke$^{1}$,
David Grellscheid$^{2}$,
Simon Pl\"atzer$^{2,3}$,
Michael Rauch$^{1}$,
\speaker{Christian Reuschle\,}$^{\,\,\,1,4}$,
Peter Richardson$^{2,5}$,
Peter Schichtel$^{2}$,
Michael H. Seymour$^{3}$,
Andrzej Si\'odmok$^{5,6}$,
Alexandra Wilcock$^{2}$,
Nadine Fischer$^{1}$,
Marco A. Harrendorf$^{7}$,
Graeme Nail$^{3}$,
Andreas Papaefstathiou$^{5}$,
Daniel Rauch$^{1}$\\
${}^{1}$Institute for Theoretical Physics, Karlsruhe Institute of Technology\\
${}^{2}$IPPP, Department of Physics, Durham University\\
${}^{3}$Particle Physics Group, School of Physics and Astronomy, University of Manchester\\
${}^{4}$HEP Theory Group, Department of Physics, Florida State University\\
${}^{5}$CERN, PH-TH, Geneva\\
${}^{6}$The Henryk Niewodniczanski Institute of Nuclear Physics, Polish Academy of Sciences\\
${}^{7}$Institut of Experimental Nuclear Physics, Karlsruhe Institute of Technology\\
}

\abstract{
With the advent and recent extension of the BLHA standard to interface Monte Carlo event generators
and one-loop matrix element providers, the Herwig++ event generator has
expanded its range of applicability to a multitude of underlying hard processes
at NLO QCD. The new NLO development is centered around the Matchbox framework,
which turns fixed NLO QCD calculations into parton shower matched calculations
- to be matched to the two parton shower variants of Herwig++. Matchbox
provides thereby for the automated setup of the underlying fixed NLO QCD
calculations and the interface to the one-loop matrix element providers, as
well as for an efficient and automated multi-channel phase space sampling, and
forms the basis for the NLO capabilities of the new release of Herwig++.
Along with several other new features and developments, the new release marks
the end of distinguishing Herwig++ and (Fortran) HERWIG, and constitutes the
first major release of version 7 of the Herwig event generator.
}

\FullConference{12th International Symposium on Radiative Corrections (Radcor 2015) and LoopFest XIV (Radiative Corrections for the LHC and Future Colliders)\\
		15-19 June  2015\\
		UCLA Department of Physics \& Astronomy Los Angeles, CA, USA}

\begin{document}

\section{Introduction}

The new release of version 7 of the Herwig event generator~\cite{Bellm:2015jjp}
marks the end of distinguishing Herwig++~\cite{Bahr:2008pv,Bellm:2013lba} and
HERWIG~\cite{Corcella:2000bw} and the point at which the physics capabilities of both
its predecessors (from the Herwig++ 2 series and the HERWIG 6 series) are
superseded.

Herwig is a computer program for the simulation of exclusive event generation
in particle collisions at hadron-hadron, hadron-lepton, and lepton-lepton
colliders. It is written in the programming language C++ and based on ThePEG,
which in the new version 2.0 is available for download along with Herwig 7.0 at
\href{https://herwig.hepforge.org/downloads}{https://herwig.hepforge.org/downloads}.

Herwig features the full simulation of particle collision events up to the
particle level, i.e. perturbative as well as non-perturbative physics. The
perturbative part provides the simulation of
hard processes at NLO QCD (including several built-in LO and NLO matrix
elements, LH event file input as well as the fully automated assembly of NLO
QCD calculations for almost all Standard Model processes, utilizing various
interfaces to several external matrix element providers),
shower Monte Carlos (facilitating two coherent shower algorithms - an
angular-ordered parton shower~\cite{Gieseke:2003rz} as well as a dipole
shower~\cite{Platzer:2009jq}, including the simulation of decays with full spin
correlations),
as well as the corresponding LO and NLO matching procedures (dedicated matrix
element corrected shower plug-ins and built-in Powheg matched cross sections,
as well as a fully automated matching machinery, with algorithms based on
MC@NLO-~\cite{Frixione:2002ik} and Powheg-type~\cite{Nason:2004rx} matching).
The non-perturbative part offers the simulation of
the hadronization process (utilizing the cluster hadronization model)
as well as of the underlying event (utilizing an eikonal multiple interaction
model).
In addition, Herwig features a highly flexible machinery for BSM
processes (including built-in BSM matrix elements as well as UFO model input
capabilities).

\section{NLO Automatization and Matching}

Automated NLO matching and merging requires full control over the fixed-order
input and hence the need for a fully integrated framework arises. The
previoulsy developed Matchbox framework~\cite{Platzer:2011bc} provides such
functionality and its extension forms the basis for the automated NLO
capabilities of the new Herwig release.

Based on the Matchbox framework the new Herwig release facilitates the
automated setup of all processes and ingredients necessary for a full NLO QCD
calculation in the subtraction formalism:
an implementation of the dipole subtraction method based on the approach by
Catani and Seymour (including massive dipole
subtraction)~\cite{Catani:1996vz,Catani:2002hc},
as well as interfaces to various external matrix element providers,
or to in-house calculations for the hard sub-processes
such as electroweak corrections to the production of heavy vector boson
pairs~\cite{Bierweiler:2013dja,Bierweiler:2012kw,Gieseke:2014gka}
or amplitudes for electroweak Higgs plus jets production provided by the
HJets++ plug-in to Matchbox.
Fully automated matching algorithms are available, inspired by
MC@NLO-~\cite{Frixione:2002ik} and Powheg-type~\cite{Nason:2004rx} matching (referred
to as subtractive and multiplicative matching respectively), for the systematic
and consistent combination of NLO QCD calculations with both shower variants in
Herwig~\cite{Gieseke:2003rz,Platzer:2009jq}.

Consider the subtractive matching for example. An observable $O_\mathrm{NLO}$
at NLO QCD, whose leading order contribution for $n$ final state partons is
given by $O_\mathrm{LO} = \int\mathrm{d}\Phi_n\mathpzc{B}O_n$, can in a
condensed notation in the subtraction formalism be written as
\begin{align}
O_\mathrm{NLO} =
\int\mathrm{d}\Phi_n \Big(\mathpzc{B}+\bar{\mathpzc{V}}\Big)O_n +
\int\mathrm{d}\Phi_{n+1} \Big(\mathpzc{R}O_{n+1} - \mathpzc{A}O_n\Big)
\;\;,\quad
\bar{\mathpzc{V}} = \mathpzc{V}+\int\mathrm{d}\Phi_1\mathpzc{A} 
\label{eq:ONLO}
\end{align}
where $\mathpzc{B}$ denotes the corresponding Born matrix element,
$\mathpzc{V}$ the UV-subtracted virtual contribution, $\mathpzc{R}$ the real
contribution and $\mathpzc{A}$ the subtraction term (the contribution from the
collinear counter term to subtract additional initial state collinear
singularities is hereby neglected). The subtractive matching formula can
correspondingly be written in a condensed notation as shown in
eq.~(\ref{eq:MCatNLO})
\begin{figure}[h]
\vspace{1.25cm}
\begin{align}
O^{\rm{matched}}_{\rm{NLO}} = 
 \Bigg[ &
          \displaystyle\int\mathrm{d}\Phi_n\Big(\tikzmark[firstA]{\mathpzc{B}}+\tikzmark[firstB]{\bar{\mathpzc{V}}}\Big)
          +
          \displaystyle\int\mathrm{d}\Phi_n\tikzmark[secondA]{\mathpzc{B}}\displaystyle\int^{Q^2}_{\mu^2_{IR}}\mathrm{d}\tikzmark[secondB]{\!\!\!\mathpzc{P}(q^2)}
          -
          \displaystyle\int\mathrm{d}\Phi_{n+1}\tikzmark[third]{\mathpzc{A}}\Theta(q^2-\mu^2_{IR})
 \Bigg] O_n \notag\\
 +
 \Bigg[ &
          \displaystyle\int\mathrm{d}\Phi_{n+1}\tikzmark[fourth]{\mathpzc{R}}
          -
          \displaystyle\int\mathrm{d}\Phi_n\tikzmark[fifthA]{\mathpzc{B}}\displaystyle\int^{Q^2}_{\mu^2_{IR}}\mathrm{d}\tikzmark[fifthB]{\!\!\!\mathpzc{P}\big(q^2\big)}
          -
          \displaystyle\int\mathrm{d}\Phi_{n+1}\tikzmark[sixth]{\mathpzc{A}}\Theta\big(\mu^2_{IR}-q^2\big)
 \Bigg] O_{n+1}
\label{eq:MCatNLO}
\end{align}
\begin{tikzpicture}[remember picture, overlay]
\draw (3,4) node[font=\footnotesize,draw=red,rounded corners,fill=red!10] (description1)
{{}$\begin{matrix}
    \textcolor{green}{A_n^{(0)},A_n^{(1)}}\\
    \textcolor{blue}{|A_n^{(0)}|^2,\langle A_n^{(0)}|A_n^{(1)}\rangle,|A_n^{(0)}\!|^2_{ij}}
    \end{matrix}$};
\draw[->] (description1) -- (firstA);
\draw[->] (description1) -- (firstB);
\draw[->] (description1) -- (secondA);
\draw[->] (description1) -- (fifthA);
\end{tikzpicture}
\begin{tikzpicture}[remember picture, overlay]
\draw (8.5,4) node[font=\footnotesize,draw=red,rounded corners,fill=red!10] (description2)
{{}$\begin{matrix}
    \textcolor{red}{P(\tilde{q}),D(p_{\!\bot})}\\
    \textcolor{red}{R_{MEC}(p_{\!\bot})}
    \end{matrix}$};
\draw[->] (description2) -- (secondB);
\end{tikzpicture}
\begin{tikzpicture}[remember picture, overlay]
\draw (12,4) node[font=\footnotesize,draw=red,rounded corners,fill=red!10] (description3)
{{}$\begin{matrix}
    \textcolor{green}{A_n^{(0)}}\\
    \textcolor{blue}{|A_n^{(0)}|^2_{ij}}
    \end{matrix}$};
\draw[->] (description3) -- (third);
\end{tikzpicture}
\begin{tikzpicture}[remember picture, overlay]
\draw (3.5,-0.75) node[font=\footnotesize,draw=red,rounded corners,fill=red!10] (description4)
{{}$\begin{matrix}
    \textcolor{green}{A_{n+1}^{(0)}}\\
    \textcolor{blue}{|A_{n+1}^{(0)}|^2}
    \end{matrix}$};
\draw[->] (description4) -- (fourth);
\end{tikzpicture}
\begin{tikzpicture}[remember picture, overlay]
\draw (7,-0.75) node[font=\footnotesize,draw=red,rounded corners,fill=red!10] (description5)
{{}$\begin{matrix}
    \textcolor{red}{P(\tilde{q}),D(p_{\!\bot})}\\
    \textcolor{red}{R_{MEC}(p_{\!\bot})}
    \end{matrix}$};
\draw[->] (description5) -- (fifthB);
\end{tikzpicture}
\begin{tikzpicture}[remember picture, overlay]
\draw (10.75,-0.75) node[font=\footnotesize,draw=red,rounded corners,fill=red!10] (description6)
{{}$\begin{matrix}
    \textcolor{green}{A_n^{(0)}}\\
    \textcolor{blue}{|A_n^{(0)}|^2_{ij}}
    \end{matrix}$};
\draw[->] (description6) -- (sixth);
\end{tikzpicture}
\vspace{1.25cm}
\end{figure}

\noindent where $\mathpzc{P}(q^2)$ denotes the shower kernel, evaluated at the
shower evolution scale $q$, $Q$ the hard process scale (aka the shower start
scale) and $\mu_\mathrm{IR}$ the shower cut-off scale, and which is rearranged
with respect to $O_n$ and $O_{n+1}$, i.e. in terms of so-called $S$- and
$H$-events respectively.

For the Born, virtual, real and subtraction term contributions various
interfaces to matrix element code exist:
either at the level of squared matrix elements through the BLHA interface
standard~\cite{Binoth:2010xt,Alioli:2013nda,Andersen:2014efa} to various
external matrix element providers such as GoSam~\cite{Cullen:2014yla},
NJet~\cite{Badger:2012pg}, OpenLoops~\cite{Cascioli:2011va} or
VBFNLO~\cite{Arnold:2008rz,Baglio:2014uba},
or at the level of color ordered sub-amplitudes with a dedicated interface to
MadGraph~\cite{Alwall:2011uj,Alwall:2014hca}, where the color bases are
provided by an interface to the ColorFull~\cite{Sjodahl:2014opa} and
CVolver~\cite{Platzer:2013fha} libraries, or at the level of built-in helicity
sub-amplitudes, with dedicated spinor-helicity libraries and caching
facilities.
For the shower subtraction plug-ins for the shower algorithms of Herwig are
provided: for the angular-ordered parton shower ($P(\tilde{q})$), the dipole
shower ($D(p_{\!\bot})$), or in the form of matrix element corrections.

For the multiplicative matching, additional capabilities to sample Sudakov-type
distributions are provided~\cite{Platzer:2011dr}.

\section{Shower Improvements, Uncertainties and Tuning}

The angular-ordered parton shower of the new Herwig release finally reaches
the same level of accuracy as in HERWIG 6. This is due to a number of
improvements.
The inclusion of QED radiation: the maximum scale for QED radiation differs in
general from the maximum scale for QCD radiation and is selected from the other
charged particles in the process rather than from colored partners. Which type
of emission is generated first is decided according to a competition algorithm:
trial QCD and QED emissions are generated and the one with the higher scale is
chosen. Depending on which type of emission suceeds over the other any
subsequent emissions of the same type are required to be angular ordered, while
the others are just required to be ordered in the evolution variable.
Spin correlations: effects of two types of correlations between the azimuthal
angle of a branching and both the hard scattering process and any previous
branchings in the parton shower are included, using the algorithm
of~\cite{Knowles:1988vs,Knowles:1988hu,Collins:1987cp}. There is now no
requirement anymore that unstable decays are generated before the parton shower
in order to generate the spin correlations between the production and decay of
the particles, as described in~\cite{Richardson:2001df,Gigg:2007cr}, since
the decays of unstable fundamental particles are now handled as part of the
parton shower including all the spin correlations.
Relaxed conditions for $g\rightarrow q\bar{q}$ branchings: the branching
$g\to q\bar{q}$ does not have a soft singularity and therefore should not be
angular-ordered in the parton shower (but must continue to be ordered in the
evolution variable). In any $g\to q\bar{q}$ branching the maximum scale is
thus determined by the scale at which the gluon itself was produced.

Uncertainties in fixed-order only runs can be assessed by variations of
renormalization and factorization scale. Similarly in runs with subsequent
showering, variations of the renormalization and factorization scales in the
shower as well as the hard process scale for the matching can be performed.
Within the new steering formalism default settings are provided for a coherent
assessment of those uncertainties.

Due to the new developments in the shower algorithms the need for a new tune to
$e^+e^-$ data arises. This tune has been carried through and a reasonable
description of the data has been achieved for both shower variants in Herwig. A
good description of underlying event data and double parton scattering data has
been obtained by including the latter with sufficently large weight in the fit.
Herwig 7.0 is released with the tune H7-UE-MMHT, using the MMHT2014 LO PDF
set~\cite{Harland-Lang:2014zoa} (tunes using the CT14~\cite{Dulat:2015mca} and
NNPDF3.0~\cite{Ball:2014uwa} PDF sets are planned for the near future as well),
for which more details can be found on
\href{https://herwig.hepforge.org/tutorials/mpi/tunes.html}
{https://herwig.hepforge.org/tutorials/mpi/tunes.html}
or in~\cite{Bellm:2015jjp}.

\section{Availability and Usage}

In the new Herwig release NLO event simulation is now possible without the
requirement of separately running external codes and/or dealing with
intermediate event sample files.
Slight changes have been made to improve Herwig's steering at the level of
input files, and significant improvements are provided to integration and
unweighting, including parallelization to meet the requirements of more
complex processes.
By virtue of the Matchbox framework the new release introduces some new run
modes in compliance to two new integrator modules which provide far superior
performance compared to the old default, ACDC of ThePEG, especially for more
complex processes (one of them is based on the standard sampling algorithm
contained in the ExSample library~\cite{Platzer:2011dr}, the other is based on
the MONACO algorithm, a VEGAS~\cite{Lepage:1980dq} variant, used by
VBFNLO~\cite{Arnold:2008rz,Baglio:2014uba}).
Both integrators require an integration grid to be set up prior to the event
generation, and hence two levels of run mode have been introduced in addition
to the old \texttt{read} and \texttt{run} steps.
The new \texttt{integrate} step performs the grid adaptation, and can be
trivially parallelized to be submitted to any standard batch or grid queues.
The event generation (the \texttt{run} step) itself can of course also be
parallelized similarly.
The \texttt{integrate} step is to be preceded by a \texttt{build} step, which
will assemble the full fixed-order or matched cross section, including
subtraction terms and the possibility of external amplitude libraries
generating dedicated code for the process of interest. The old \texttt{read}
step is still available, representing the subsequent execution of the
\texttt{build} and \texttt{integrate} steps. Various examples of how this new
work-flow paradigm can be utilized are given in the new online documentation
on \href{https://herwig.hepforge.org/tutorials/index.html}
{https://herwig.hepforge.org/tutorials/index.html}.

A variety of related project code has been developed along with Herwig 7, which
is not supported at the same level as the core Herwig code, but is nevertheless
provided along with it in the Herwig \texttt{Contrib} folder. Plug-ins to
various projects exist, such as electroweak Higgs plus jets
production~\cite{Campanario:2013fsa}, provided by the HJets++ plug-in to Matchbox,
FxFx merging~\cite{Frederix:2012ps} for $\mathrm{W+jets}$ and $\mathrm{Z+jets}$
events~\cite{Frederix:2015eii} or Higgs boson boson pair
production~\cite{Goertz:2014qta,Maierhofer:2013sha}. More details can be found
in~\cite{Bellm:2013lba,Bellm:2015jjp} and in the Contrib section on
\href{https://herwig.hepforge.org/tutorials/index.html}
{https://herwig.hepforge.org/tutorials/index.html}.

\section{Examples}

\subsection{Checks and Validation}

\begin{figure}[h]
\centering
\vspace{1.5ex}
\begin{minipage}{.45\textwidth}
  \centering
  \includegraphics[trim=140 452.5 240 140, clip, scale=0.625]{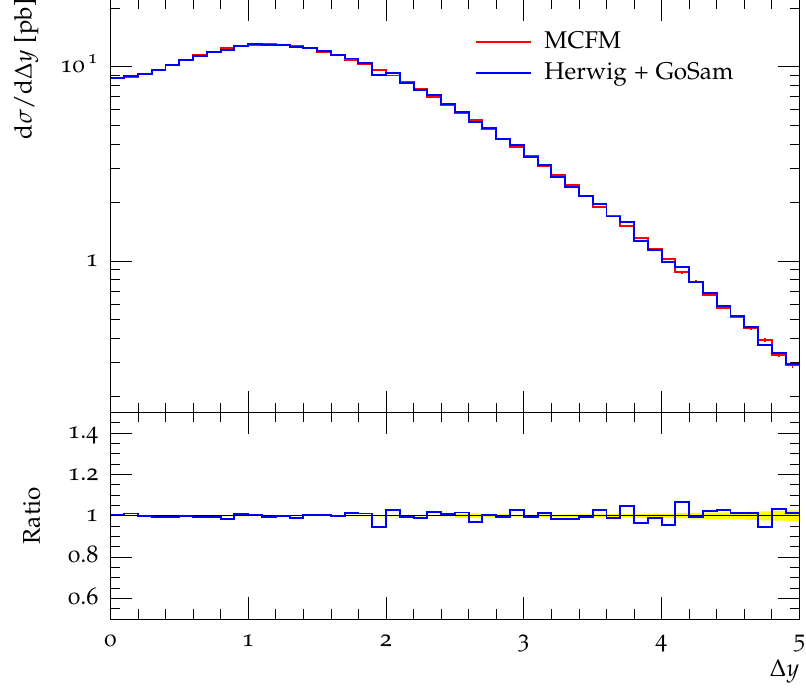}
  \caption{\small\it Validation against MCFM for $WZ$ production: $\Delta y$ between $W$ and $Z$ boson.}
  \label{fig:test3}
\end{minipage}%
\hspace{0.05\textwidth}
\begin{minipage}{.45\textwidth}
  \centering
  \hspace{-1.75ex}
  \includegraphics[trim=135 452.5 245 140, clip, scale=0.625]{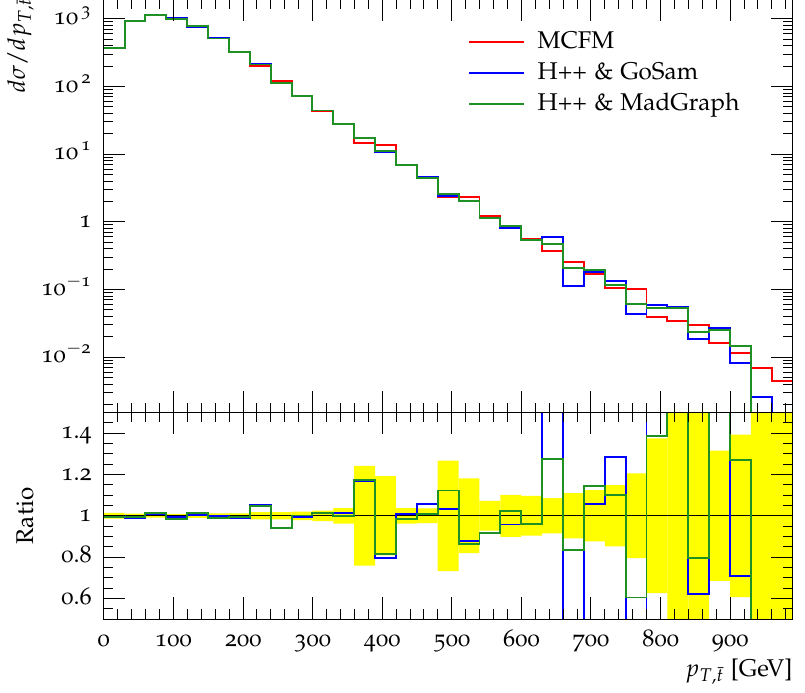}
  \caption{\small\it Validation against MCFM for $t\bar{t}$ production: $p_{T}$ of the top antiquark.}
  \label{fig:test4}
\end{minipage}
\end{figure}

\begin{figure}[h]
\centering
\vspace{-0.5ex}
\begin{minipage}{.45\textwidth}
  \centering
  \vspace{1.5ex}
  \includegraphics[trim=140 512.5 240 135, clip, scale=0.625]{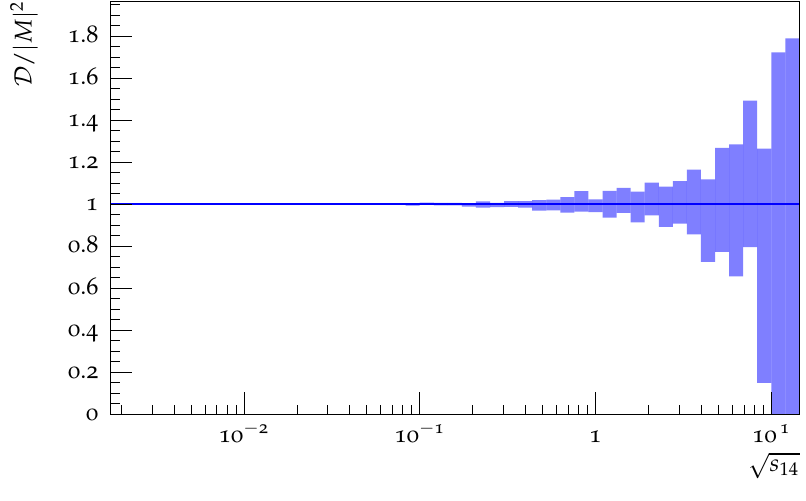}
  \caption{\small\it Real emission subtraction check for $Z+$jet production: singularity cancellation in $u\bar{u}\rightarrow e^+e^-gg$.}
  \label{fig:test5}
\end{minipage}%
\hspace{0.05\textwidth}
\begin{minipage}{.45\textwidth}
  \centering
  \vspace{-1.5ex}
  \hspace{-1.75ex}
  \includegraphics[trim=135 512.5 245 135, clip, scale=0.625]{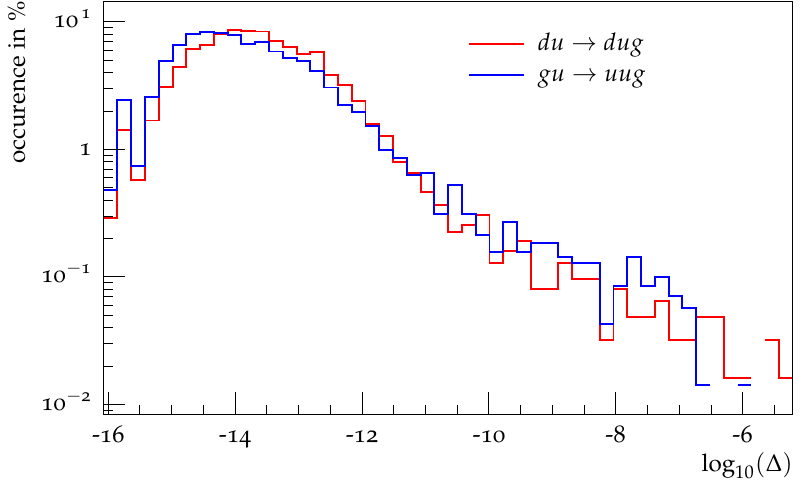}
  \caption{\small\it Pole cancellation check for $3$jet production: $\frac{1}{\varepsilon}$-pole cancellation.}
  \label{fig:test6}
\end{minipage}
\end{figure}

Figs.~\ref{fig:test3} and \ref{fig:test4} show validation plots for various
matrix element providers against results obtained from MCFM.
Figs.~\ref{fig:test5} and \ref{fig:test6} demonstrate pole cancellation and
subtraction checks.

\subsection{A more concrete example: $t\bar{t}$ Production}

\begin{figure}[h]
\centering
\includegraphics[trim=0 0 0 22.5, clip, scale=0.65]{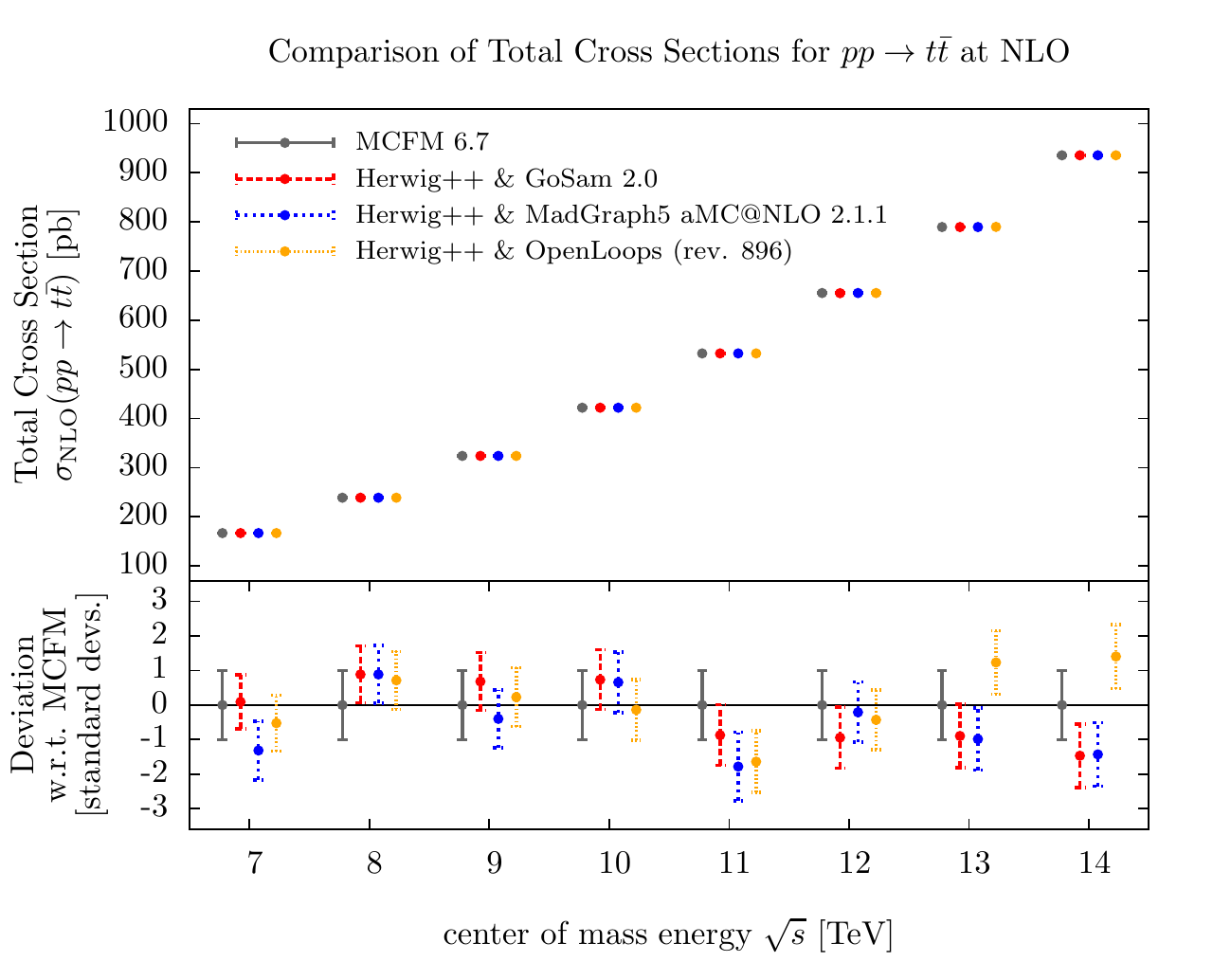}
\caption{\small\it Comparison of total cross sections for $pp\rightarrow t\bar{t}$ at NLO. Various matrix element providers against MCFM.
         Fixed scale $\mu=\mu_F=\mu_R=80$ GeV. CT10nlo PDF set, $\alpha_s(M_Z)|_{\rm{CT10nlo}}$. $M_t=173.5$ GeV, on-shell.}
\label{fig:test9}
\end{figure}

\begin{figure}[h]
\centering
\includegraphics[scale=0.325]{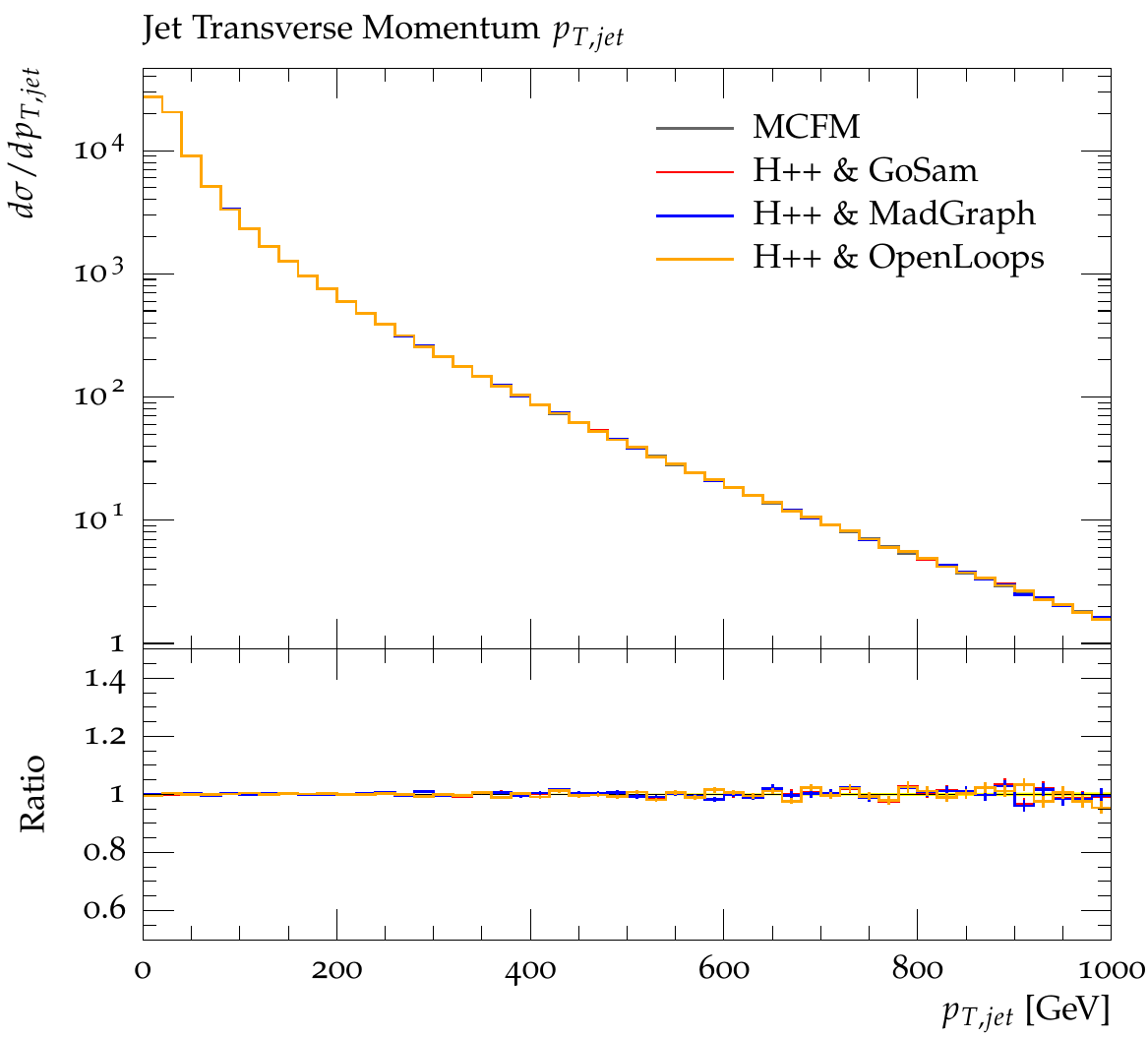}
\includegraphics[scale=0.325]{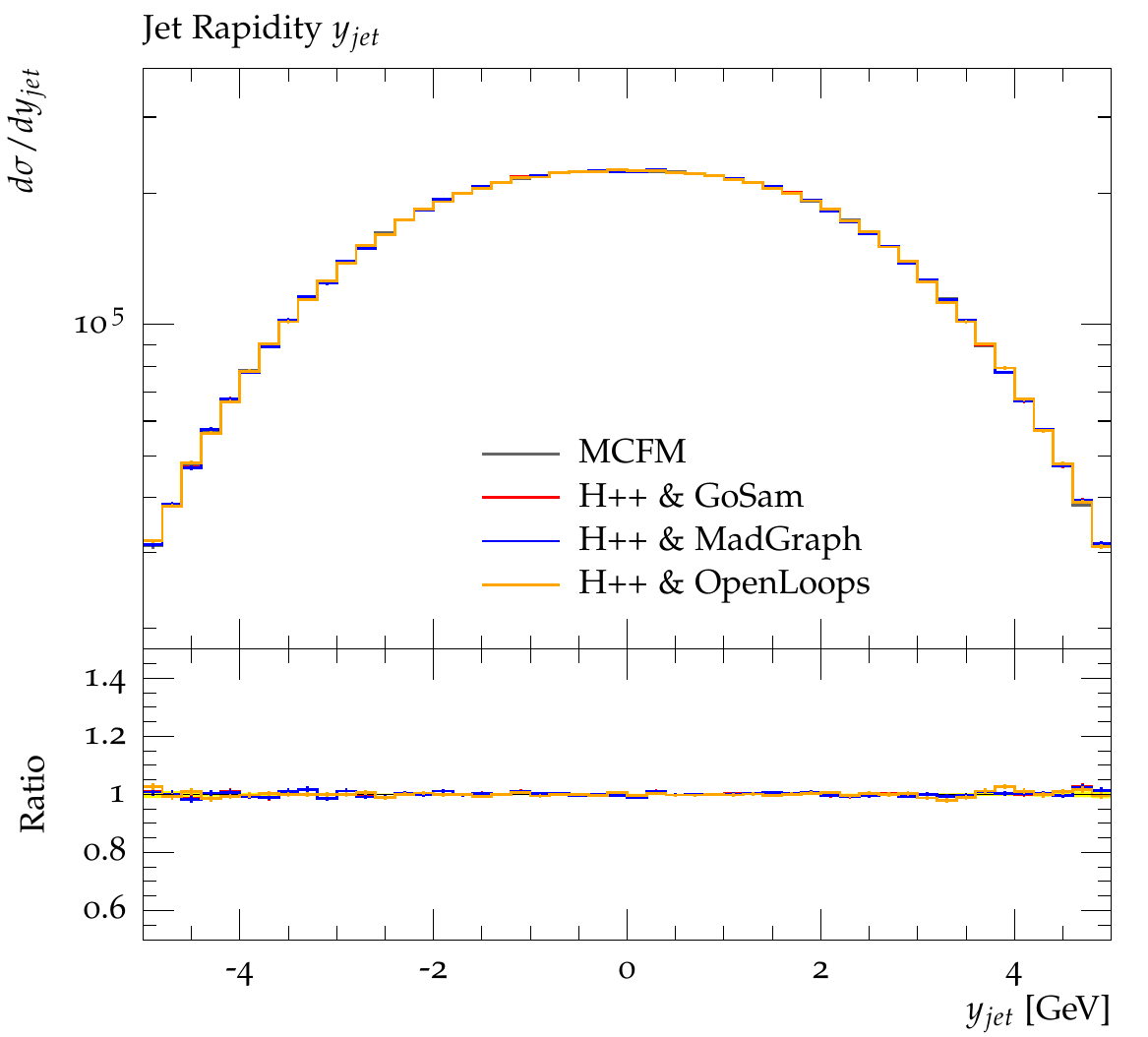}
\includegraphics[scale=0.325]{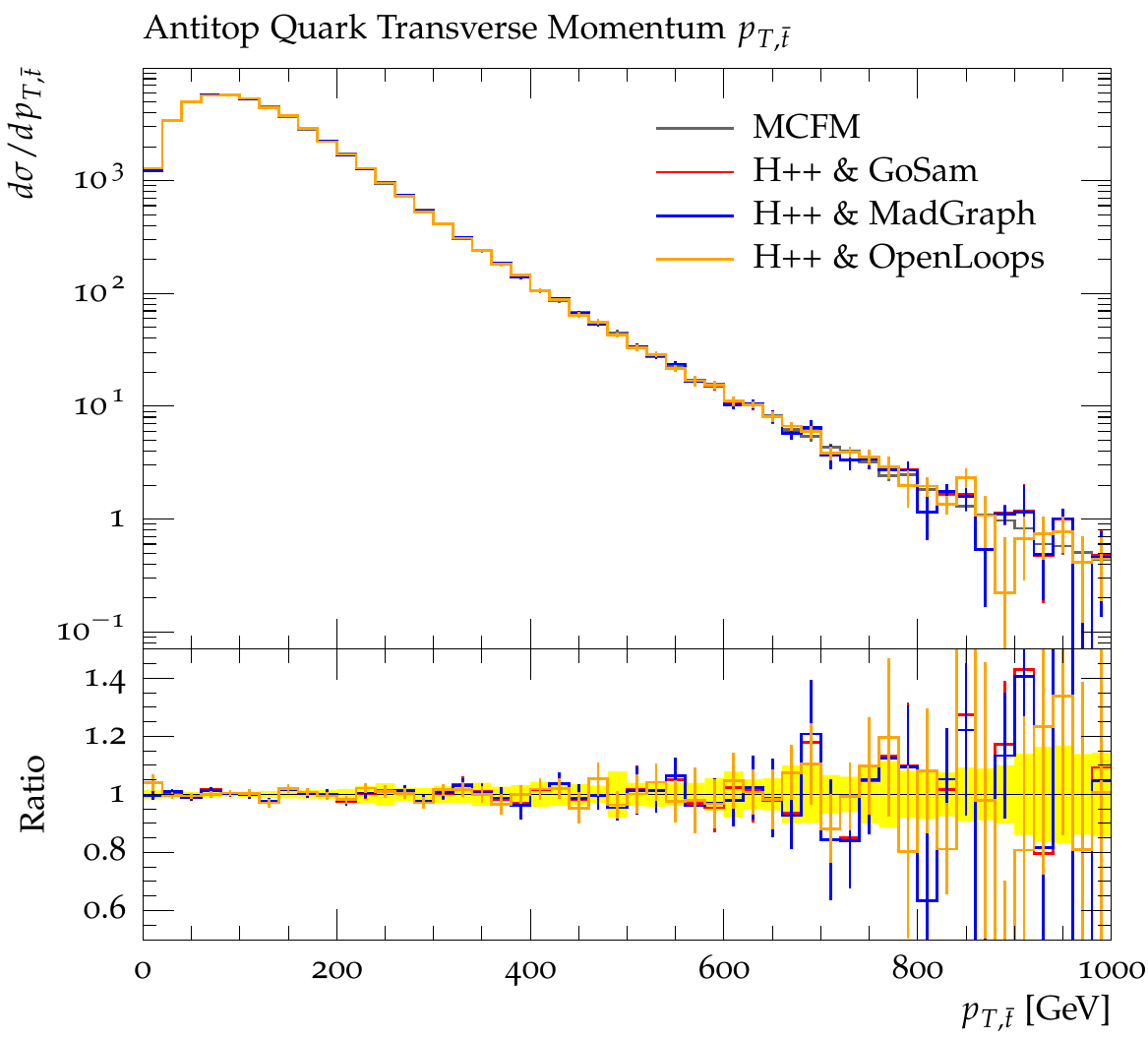}\\
\includegraphics[scale=0.325]{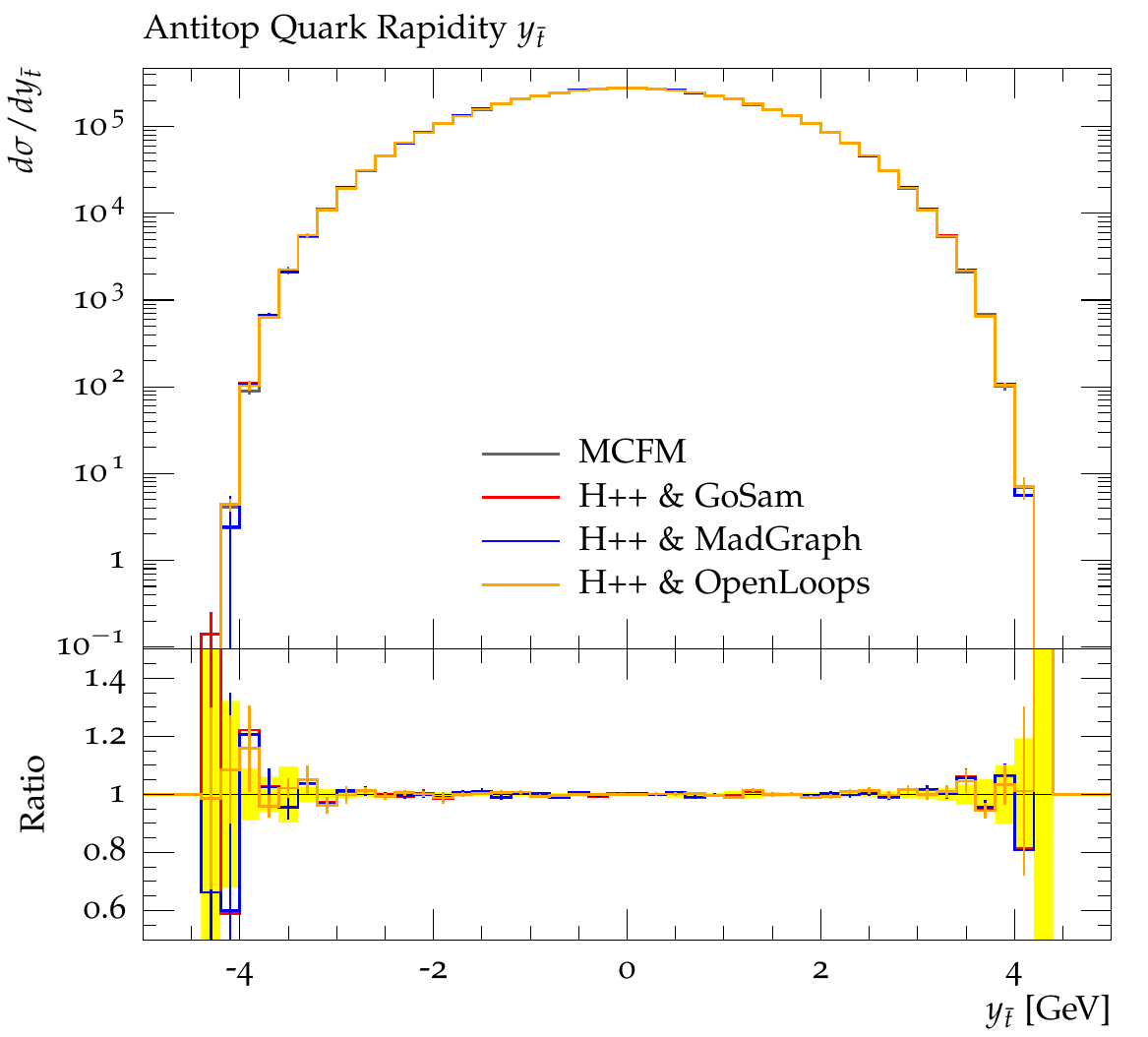}
\includegraphics[scale=0.325]{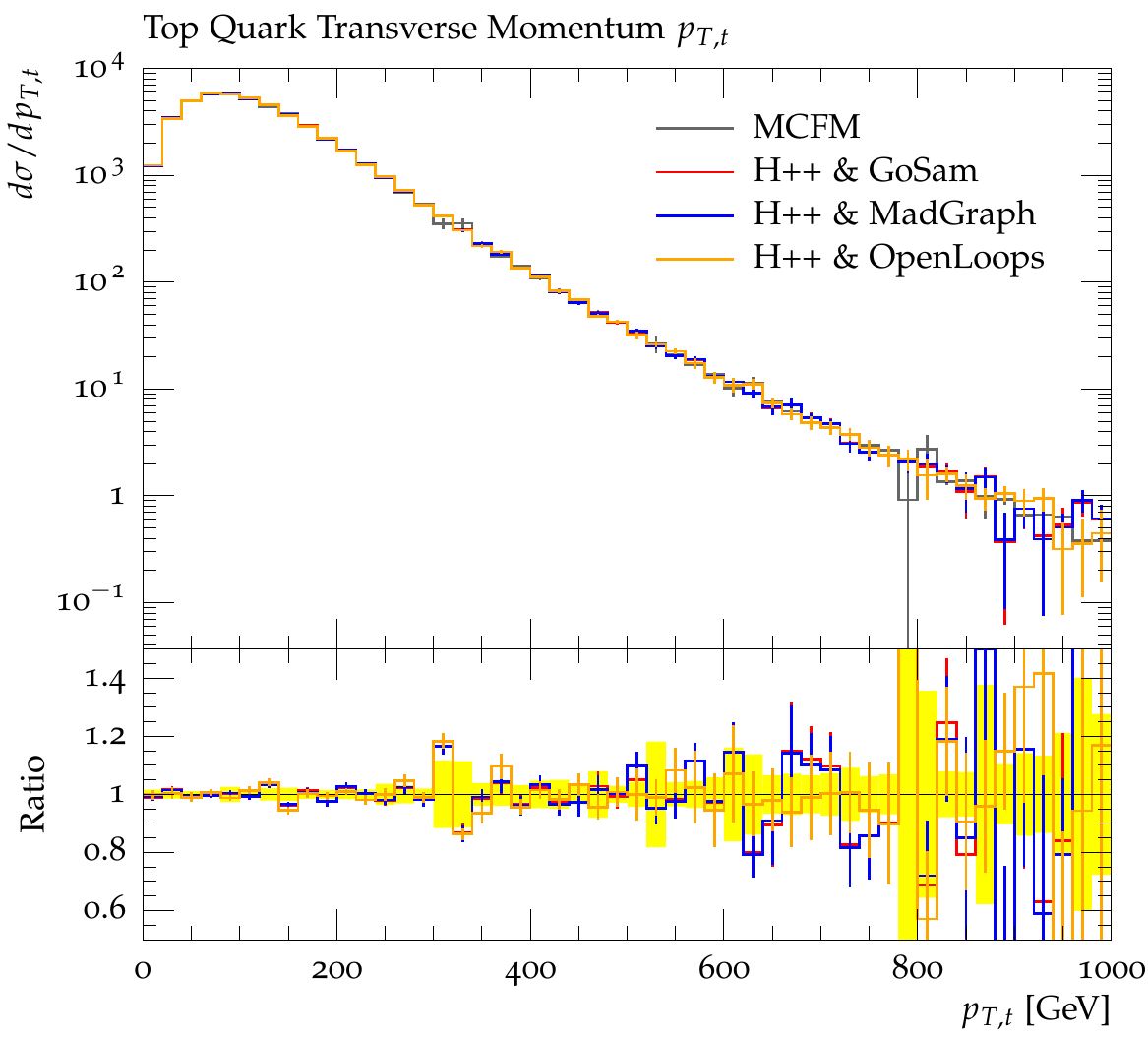}
\includegraphics[scale=0.325]{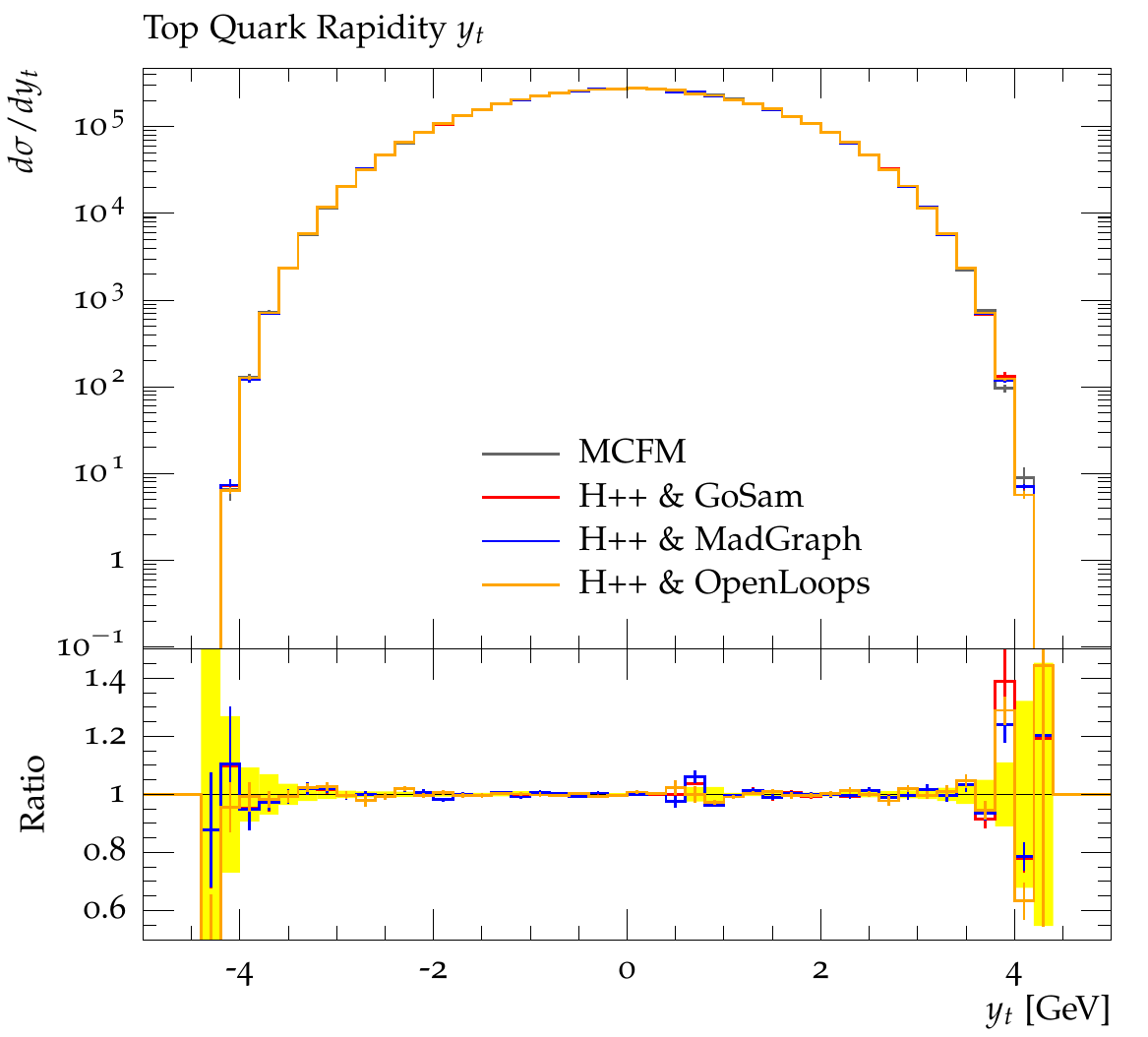}
\caption{\small\it Comparison of various distributions for $pp\rightarrow t\bar{t}$ at NLO, at $14$ TeV. Various matrix element providers against MCFM.
         Fixed scale $\mu=\mu_F=\mu_R=80$ GeV. CT10nlo PDF set, $\alpha_s(M_Z)|_{\rm{CT10nlo}}$. $M_t=173.5$ GeV, on-shell.}
\label{fig:test10}
\end{figure}

\begin{figure}[h]
\centering
\includegraphics[scale=0.5]{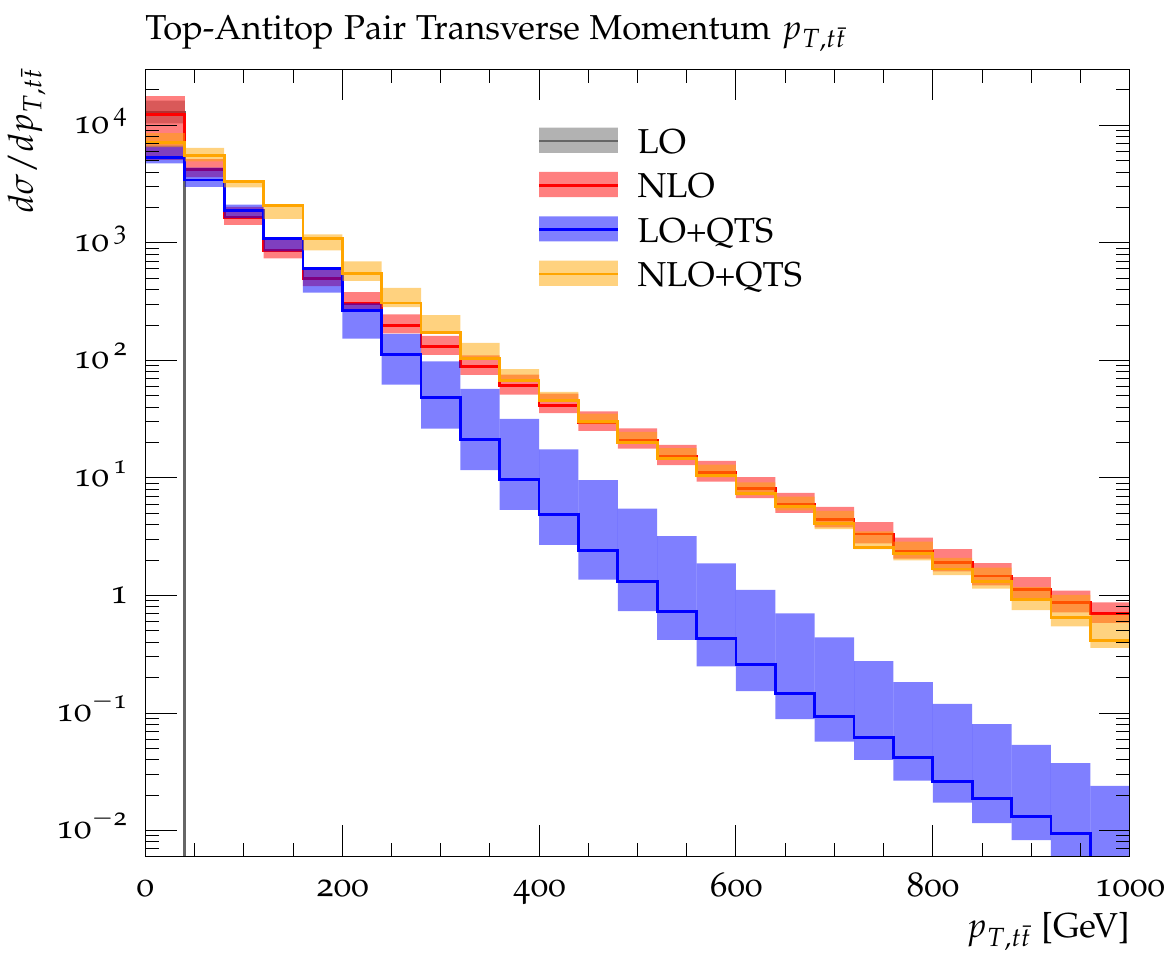}
\hspace{7.5ex}
\includegraphics[scale=0.5]{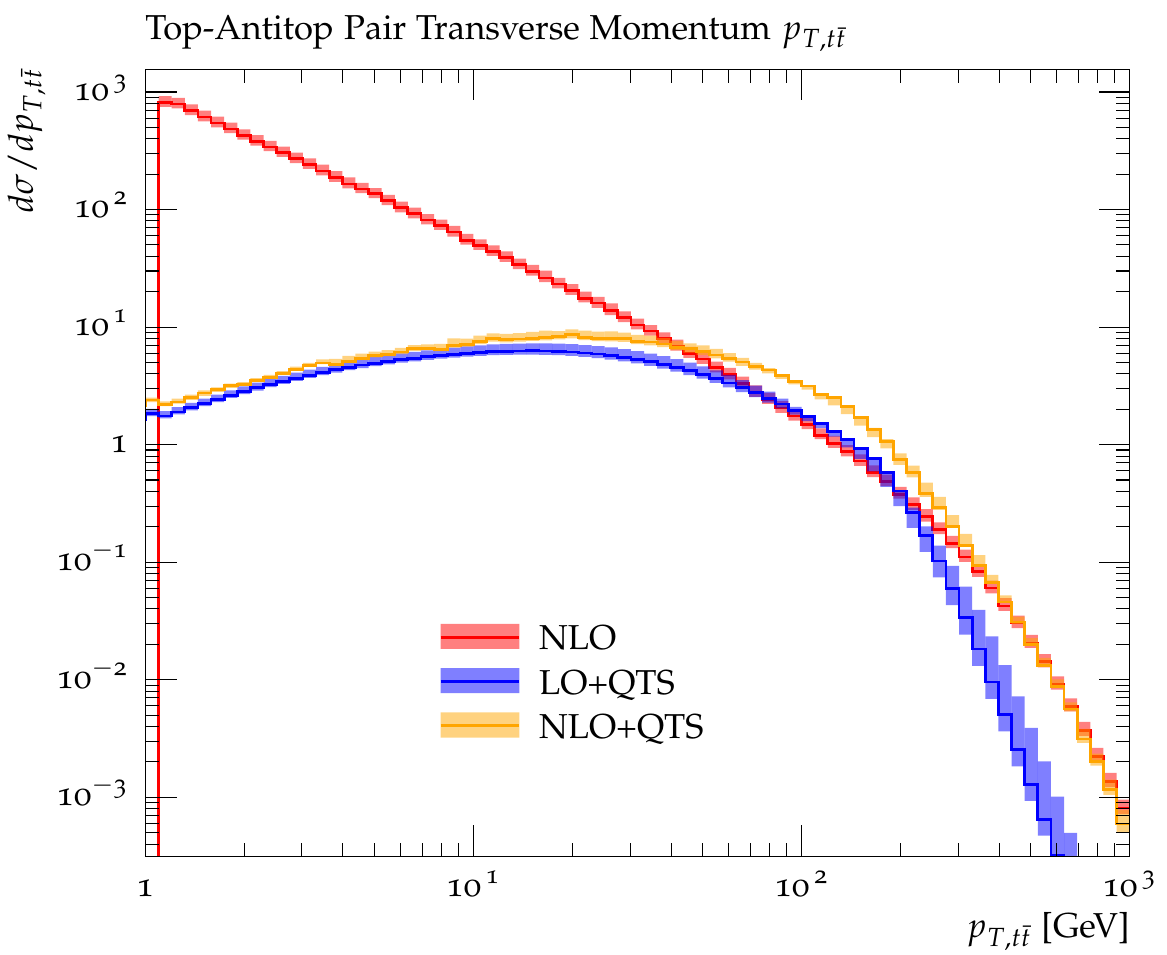}
\caption{\small\it $pp\rightarrow t\bar{t}$ NLO matched, at $14$ TeV.
         MC@NLO--like matching to the angular ordered $\tilde{q}$--shower. Shower start scale $Q^2=\mu^2$.
         Hard scale variation: $\mu^2=\mu_F^2=\mu_R^2$ by factors of $0.5$ and $2$,
         where $\mu^2=\mu_{\min_{t\bar{t}}\{m_T\}}^2 = \min_{i=t,\bar{t}} \left(m_i^2+p_{T,i}^2\right)$.
}
\label{fig:test11}
\end{figure}

Figs.~\ref{fig:test9} and \ref{fig:test10} show validation plots for various
matrix element providers against results obtained from MCFM, for $t\bar{t}$
production at NLO.
Fig.~\ref{fig:test11} shows results for $t\bar{t}$ production at NLO,
supplemented with the angular-ordered shower of Herwig and subtractive
matching.

\subsection{Data Comparison and Comparison vs. Herwig++ 2.7}

\begin{figure}[h]
\centering
\vspace{-2.5ex}
\begin{minipage}{.45\textwidth}
  \centering
\includegraphics[scale=0.5]{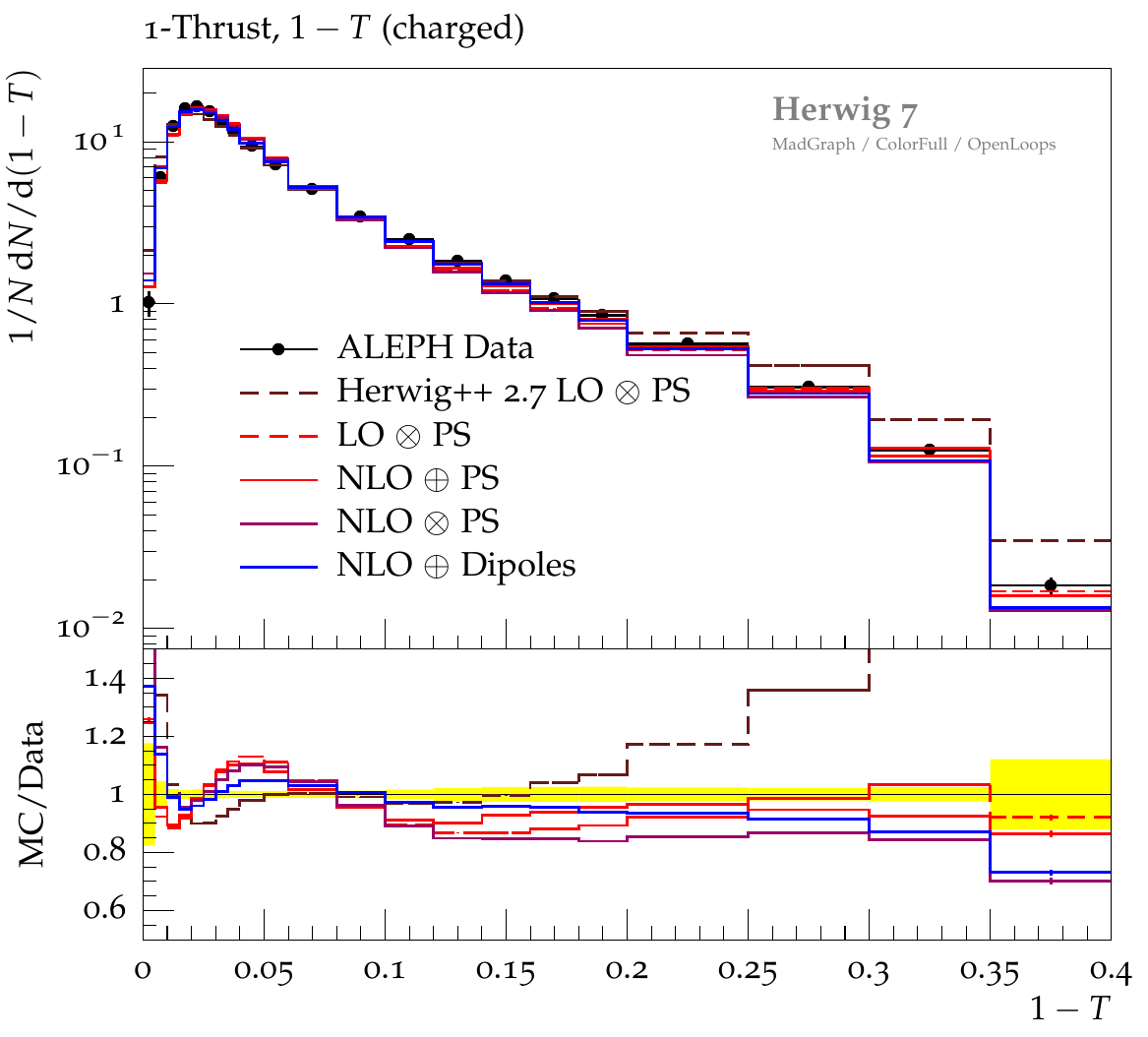}
  \caption{\small\it The thrust distribution in $\mathrm{e^+e^-}$ annihilation at
    $\sqrt{s}=M_z$, in comparison with ALEPH data~\cite{Barate:1996fi}.}
  \label{results1}
\end{minipage}%
\hspace{0.05\textwidth}
\begin{minipage}{.45\textwidth}
  \centering
\vspace{2.5ex}
\includegraphics[scale=0.5]{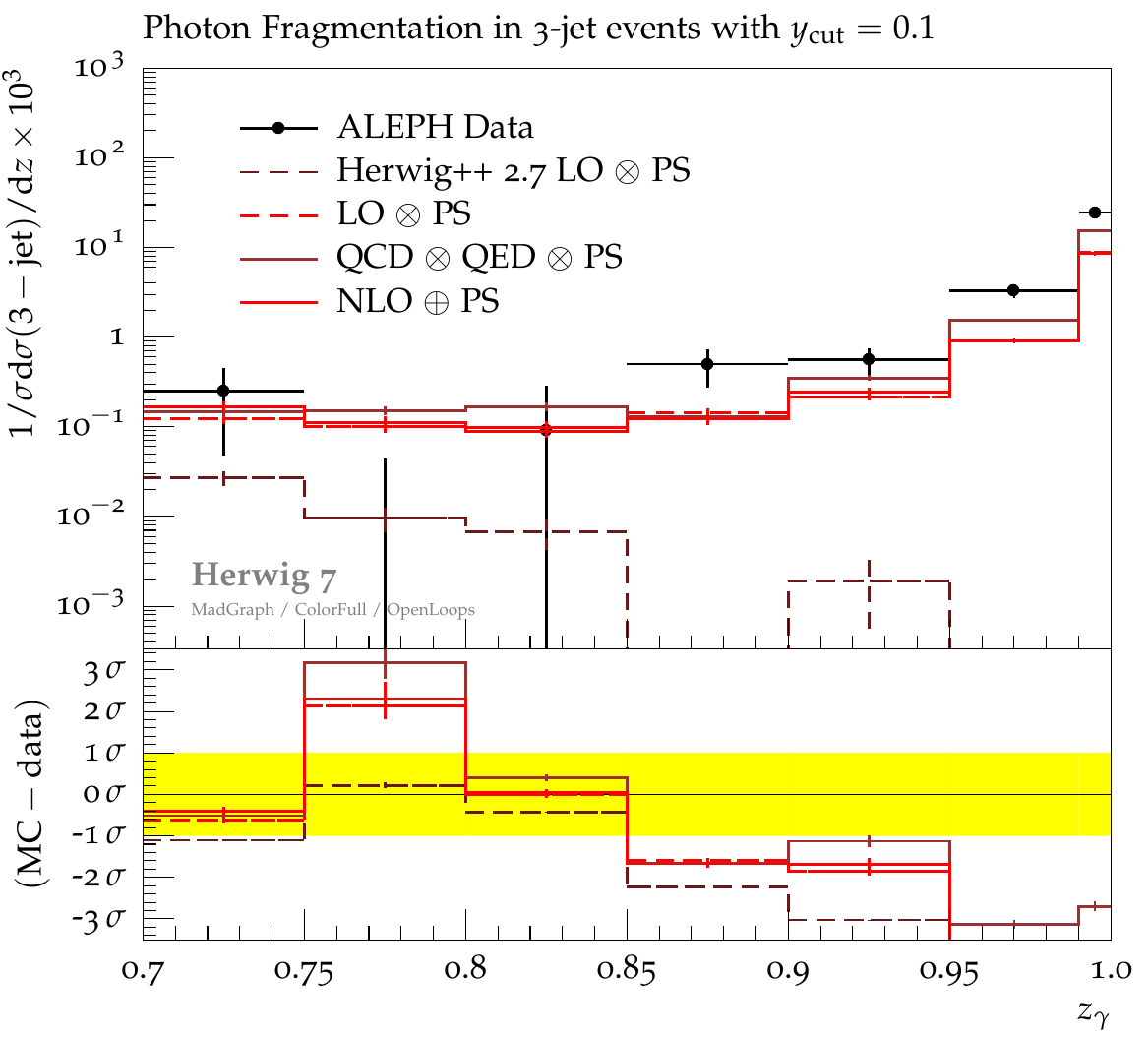}
  \caption{\small\it The distribution of photon-jet energy fraction in three-jet
    $\mathrm{e^+e^-}$ events at $\sqrt{s}=M_z$ defined with a cutoff in
    the $k_\perp$ algorithm of $y=0.1$ in comparison with ALEPH
    data~\cite{Buskulic:1995au}.}
  \label{results2}
\end{minipage}
\end{figure}

\begin{figure}[h]
\centering
\vspace{-2.5ex}
\begin{minipage}{.45\textwidth}
  \centering
\includegraphics[scale=0.5]{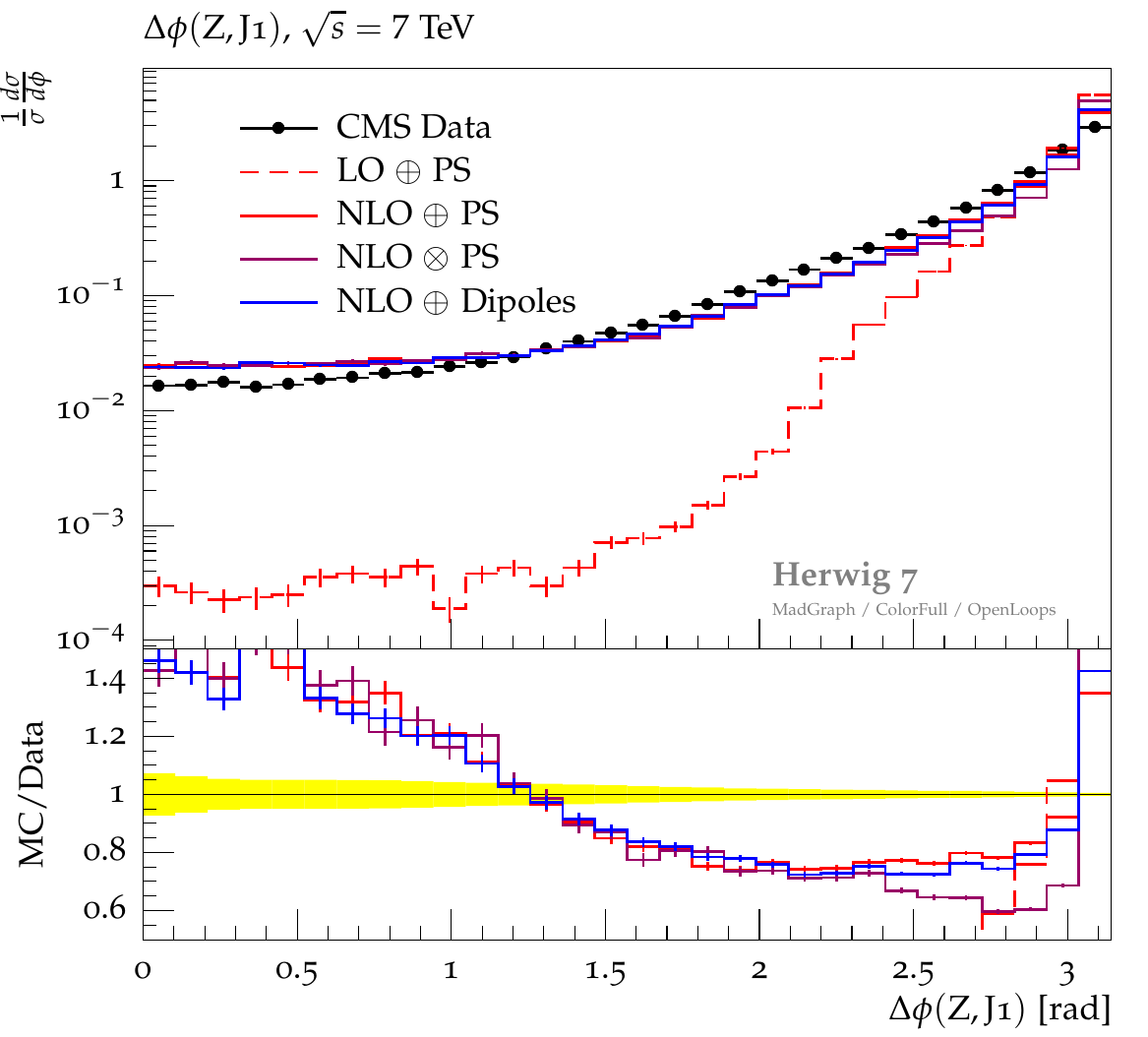}
  \caption{\small\it The distribution of separation in azimuthal angle between
    the $\mathrm{Z}$ boson and the hardest jet in $\mathrm{Z+jets}$
    events in $\mathrm{pp}$ collisions at $\sqrt{s}=7\,\mathrm{TeV}$ in
    comparison with CMS data~\cite{Chatrchyan:2013tna}.}
  \label{results3}
\end{minipage}%
\hspace{0.05\textwidth}
\begin{minipage}{.45\textwidth}
  \centering
\vspace{2.5ex}
\includegraphics[scale=0.5]{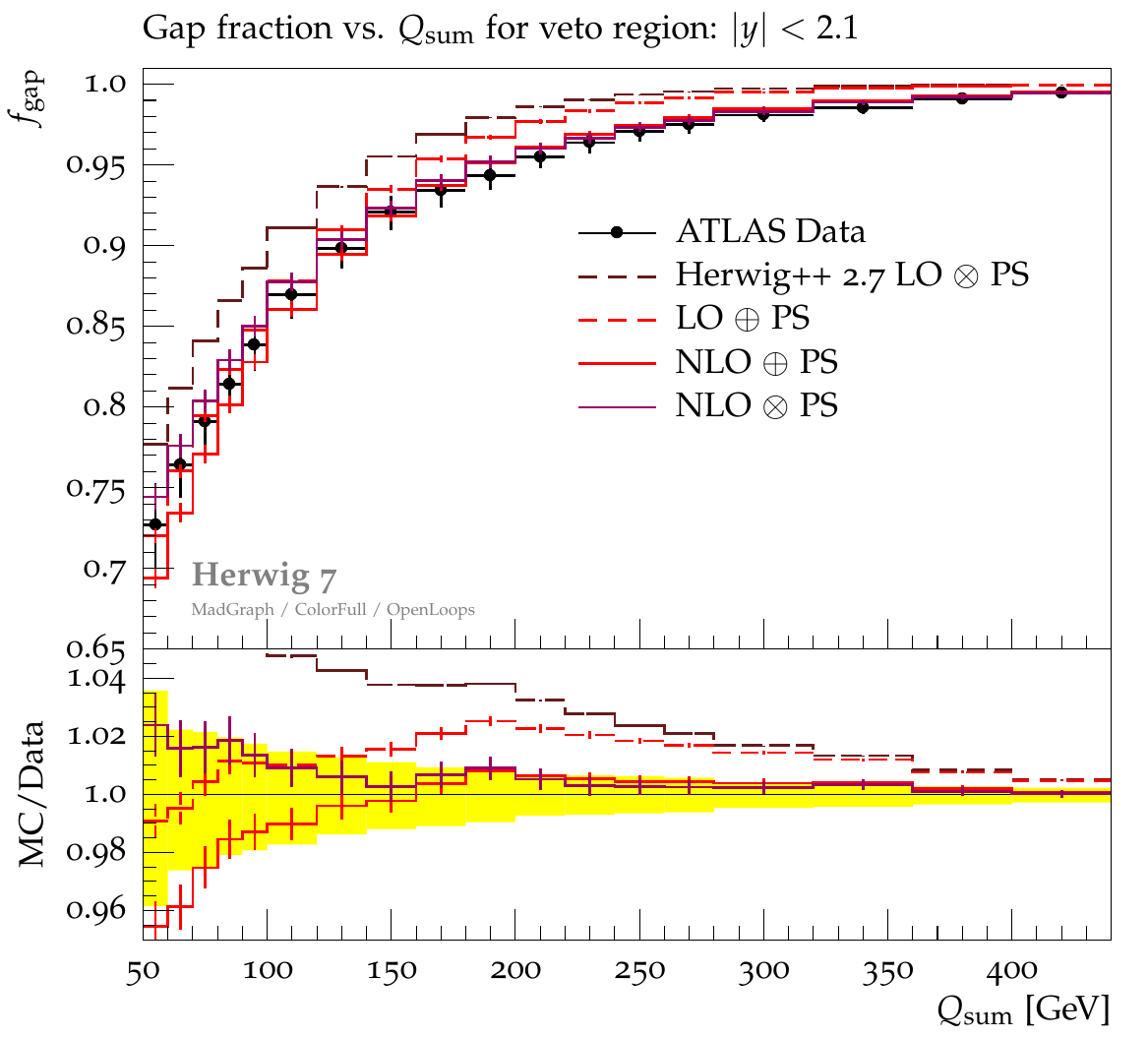}
  \caption{\small\it The fraction of events that have less than $Q_{\mathrm{sum}}$
    transverse energy in the rapidity region $|y|<2.1$ in top
    quark-antiquark events in $\mathrm{pp}$ collisions at
    $\sqrt{s}=7\,\mathrm{TeV}$ in comparison with ATLAS
    data~\cite{ATLAS:2012al}.}
  \label{results4}
\end{minipage}
\vspace{-1.0ex}
\end{figure}

In figs.~\ref{results1} to \ref{results4} a small sample of results is shown
that have been improved in the new release of Herwig 7.0 in contrast to the
previous Herwig++ version 2.7. The following is just a small selection of
distributions that have been checked against data. More will be shown in a
series of upcoming papers.
The Monte Carlo results from Herwig++ 2.7 use leading order plus parton
shower,
those from Herwig 7.0 use
the angular-ordered parton shower (LO $\oplus$ PS),
the angular-ordered parton shower supplemented
  by the built-in Powheg correction, which includes QCD and QED corrections for
  the case of $e^+e^-\to q\bar{q}$ (QCD $\otimes$ QED $\otimes$PS),
  by the automatically-calculated by Matchbox subtractive (MC@NLO-type)
  matching (NLO $\oplus$ PS) and 
  multiplicative (Powheg-type, NLO $\otimes$ PS) corrections
and, finally, the dipole shower supplemented by a subtractive matching to NLO
cross sections (NLO $\oplus$ Dipoles).

Fig.~\ref{results1} shows the thrust distribution at LEP, in comparison with
data from ALEPH~\cite{Barate:1996fi}. A long-standing problem of Herwig++
(producing too many very hard events, whether or not NLO matching was used)
has been been solved by the improvements to the angular-ordered parton shower:
all of the variants of NLO matching give a similar description of the data.
In Fig.~\ref{results2}, the effect of the inclusion of photon emission in the
angular-ordered parton shower is shown. Events at $z_\gamma=1$ are isolated
photons (``jets'' for which all of the jet energy is carried by a single
photon), while events at lower $z_\gamma$ come from hard collinear photon
emission from the final state quark jets. We see clearly that the results from
Herwig++ have no component at large $z_\gamma$ at all, while all of the Herwig
7.0 variants are much closer to the data.

Fig.~\ref{results3} shows results for $\mathrm{Z+jets}$ production at the LHC,
more concretely the distribution of separation in azimuthal angle between the
$\mathrm{Z}$ boson and the hardest jet. The region $\Delta\phi\sim\pi$
corresponds to leading order kinematics, in which the $\mathrm{Z}$ boson gains
transverse momentum by recoiling against a single hard parton, whereas the
spectrum of events down to $\Delta\phi=0$ corresponds to events in which the
$\mathrm{Z}$ boson recoils against two or more jets. The need for NLO
corrections is clearly seen. An important cross-check of the two different
automated NLO matching schemes and the two different shower algorithms, both
using subtractive matching, can also be seen.
In Fig.~\ref{results4} the jet activity in $\mathrm{t\bar{t}}$ events at the
LHC is shown, as revealed by the gap fraction.
Herwig++ 2.7 is seen to have far too little jet activity (too many gap events).
While Herwig 7.0 with the shower alone is somewhat closer to the data at small
$Q_{\mathrm{sum}}$, a clear deficit is seen for hard jet events at high
$Q_{\mathrm{sum}}$, while both the NLO matching schemes describe the data well.

\section{Summary and Outlook}

The Herwig event generator has expanded its range of applicability to a
multitude of underlying hard processes at NLO QCD, and a new version has
recently been released~\cite{Bellm:2015jjp}. Much of the new development is due
to the Matchbox framework for fully automated NLO calculations and matching,
which, however, also triggered crucial improvements to hitherto existing
structures. The new release is the first major release in the new Herwig 7
series, which supersedes the capabilities of both its predecessors in the
Herwig++ 2~\cite{Bahr:2008pv,Bellm:2013lba} and HERWIG 6~\cite{Corcella:2000bw}
series.
Further physics related studies adressed by the new version, as well as
comparisons with other event generator programs such as
Pythia~\cite{Sjostrand:2007gs,Sjostrand:2014zea} or Sherpa~\cite{Gleisberg:2008ta},
will be covered in subsequent publications.
Future development, for which the recent release forms the basis, includes LO
and NLO multijet merging as well as the automation of NLO electroweak
corrections.

\section*{Acknowledgements}

CR is thankful to the organizers of the Radcor 2015 and LoopFest XIV joined
symposium, for the possibility to present the new developments in Herwig.
This work was supported in part by the European Union as part of the FP7 Marie
Curie Initial Training Network MCnetITN (PITN-GA-2012-315877).

All contributors, hosting institutions and funding agencies who gave invaluable
input, feedback and support for the new release of Herwig 7 are furthermore
acknowledged in~\cite{Bellm:2015jjp}.

\bibliography{Herwig}{}
\bibliographystyle{JHEP}

\end{document}